\pdfoutput=1

\documentclass[11pt,twoside,a4paper,cmspaper,final,collab]{cms-tdr}
\def\svnVersion{277864}\def\cmsCernNo{2013-037}\def\cmsCernDate{\today}\def\cmsMessage{Published in Physical Review Letters as \href{http://dx.doi.org/10.1103/PhysRevLett.114.061801}{\doi{10.1103/PhysRevLett.114.061801.}}}

\begin{document}\cmsNoteHeader{B2G-12-024}

\hyphenation{had-ron-i-za-tion}
\hyphenation{cal-or-i-me-ter}
\hyphenation{de-vices}

\RCS$Revision: 277858 $
\RCS$HeadURL: svn+ssh://svn.cern.ch/reps/tdr2/papers/B2G-12-024/trunk/B2G-12-024.tex $
\RCS$Id: B2G-12-024.tex 277858 2015-02-17 17:55:21Z lantonel $

\newcommand{\ztt}{\ensuremath{\cPZ\to\Pgt\Pgt}\xspace}
\newcommand{\zee}{\ensuremath{\cPZ\to\Pe\Pe}\xspace}
\newcommand{\zmumu}{\ensuremath{\cPZ\to\Pgm\Pgm}\xspace}
\newcommand{\zll}{\ensuremath{\cPZ\to\ell\ell}\xspace}
\newcommand{\wlnu}{\ensuremath{\PW\to\ell\Pgn\text{ + jets}}\xspace}
\ifthenelse{\boolean{cms@external}}{\providecommand{\CL}{CL\xspace}}{\providecommand{\CL}{CL\xspace}}
\ifthenelse{\boolean{cms@external}}{\providecommand{\cmsLeft}{top\xspace}}{\providecommand{\cmsLeft}{left\xspace}}
\ifthenelse{\boolean{cms@external}}{\providecommand{\cmsRight}{bottom\xspace}}{\providecommand{\cmsRight}{right\xspace}}
\cmsNoteHeader{B2G-12-024} 

\title{Search for Displaced Supersymmetry in events with an electron and a muon with large impact parameters}

\date{\today}

\abstract{
A search for new long-lived particles decaying to leptons is presented using proton-proton
collisions produced by the LHC at $\sqrt{s} = 8\TeV$. Data used for
the analysis were collected by the CMS detector and correspond to an integrated
luminosity of 19.7\fbinv.
Events are selected with an electron and muon with opposite charges that both have transverse impact parameter values between 0.02\unit{cm} and 2\cm.
The search has been designed to be sensitive to a wide range of  models with nonprompt \Pe-$\Pgm$ final states.
Limits are set on the ``displaced supersymmetry'' model,
with pair production of top squarks decaying into an \Pe-$\Pgm$ final state via R-parity-violating interactions.
The results are the most restrictive to date on this model, with the most stringent limit being obtained for a top squark lifetime corresponding to $c\tau = 2\unit{cm}$,
excluding masses below 790\GeV at 95\% confidence level.
}

\hypersetup{%
pdfauthor={CMS Collaboration},%
pdftitle={Search for ``Displaced Supersymmetry" in events with an electron and a muon with large impact parameters},%
pdfsubject={CMS},%
pdfkeywords={CMS, physics, supersymmetry}}

\maketitle 

Although the standard model (SM) has been successful at describing the known elementary particles and their interactions, it is considered to be a low-energy effective theory.
Numerous experiments have conducted searches aimed at discovering extensions to the SM.
These searches have generally assumed short particle lifetimes resulting in prompt decay signatures.
However, since many models of physics beyond the standard model (BSM) predict particles with significant
lifetimes~\cite{DisplacedSUSY,rParityViolatingSupersymmetry,SplitSUSY,StableMassiveParticles,GMSB,HiddenValleyRanch,B-L,LongLivedHiggs},
there have been some dedicated analyses that attempt to identify particles whose production vertices were displaced from the interaction region~\cite{ALEPHLLSearch,OPALLLSearch,D0LLSearch}.
Recent results from experiments at the CERN LHC that constrain conventional BSM models have considerably increased the motivation to search for long-lived BSM particles~\cite{Craig:2013cxa}.

The CMS and ATLAS Collaborations have searched for decays of long-lived BSM particles, covering a range of decay lengths.
These include searches for heavy stable charged particles~\cite{HSCP,ATLASHSCP,StoppedHSCP,ATLASStoppedHSCP,ATLASMulticharged}, 
searches using disappearing track signatures~\cite{ATLASDisappearingTracks},
and searches focusing on short decay lengths~\cite{ATLASDisplacedJets,DisplacedMuonJetVertex,DisplacedLeptons,ATLASLeptonJets,ATLASDisplacedPhotons,DisplacedPhotons1,DisplacedPhotons2}.
The search described in this Letter is distinct from these analyses in that it does not make any assumptions about the event topology
beyond the requirement that the event contain an isolated electron and isolated muon with large impact parameters and opposite charges.
It does not require, or exclude, hadronic activity or missing transverse energy (\ETslash).
It does not require, or exclude, that the reconstructed displaced tracks form a vertex.
It does not require that the displaced tracks are collimated.
In this way, the analysis has sensitivity to a wide variety of still viable BSM scenarios that produce only pairs of displaced leptons with opposite charges.

This broad applicability notwithstanding, we interpret the search results in the context of a representative model termed ``displaced supersymmetry"~\cite{DisplacedSUSY}.
This model introduces large sparticle lifetimes via small R-parity-violating (RPV) couplings~\cite{rParityViolatingSupersymmetry}.
Consequently, the resultant decay products can be measurably displaced in the transverse plane relative to the proton beams.
We consider the case in which the lightest supersymmetric particle is the top squark ($\sTop$),
which is produced in pairs and then decays through an RPV vertex,
$\sTop {\to} \cPqb \Pl$, where $\Pl$ is any charged lepton.
The search does have sensitivity to leptonic tau decays, although the acceptance is suppressed by the small branching fraction of taus to leptons and by the soft \pt spectrum of these leptons.
The production mechanism is given entirely by the QCD of massive color triplet scalars and depends only on the stop mass and not on any other supersymmetric parameters.
For simplicity we assume lepton universality in the top squark decay vertex.
We conduct a search for an electron and a muon produced from different top squarks in the final state, with both of the leptons having large impact parameters.
We restrict the search to include only tracks with impact parameter less than 2\unit{cm},
for which the CMS triggering and reconstruction algorithms have previously been optimized.

The central feature of the CMS apparatus is a
superconducting solenoid of $6\unit{m}$ internal diameter. Within the magnet
volume are a silicon pixel and strip tracker, a lead tungstate crystal
electromagnetic calorimeter (ECAL), and a brass and scintillator hadron calorimeter, each composed of a barrel and two end cap sections.  Muons are measured
in gas-ionization detectors embedded in the steel flux-return yoke outside the
solenoid. Extensive forward calorimetry complements the coverage provided by
the barrel and end cap detectors.
A more detailed description of the CMS detector,
together with a definition of the coordinate system used and the relevant kinematic variables,
can be found in Ref.~\cite{Chatrchyan:2008zzk}.

The analysis uses data from proton-proton collisions recorded at $\sqrt{s} = 8\TeV$ which
correspond to an integrated luminosity of $(19.7 \pm 0.5)\fbinv$.
Events were recorded with a trigger that requires a muon with $\pt > 22\GeV$ and a cluster in the ECAL with transverse energy $\et > 22\GeV$.
No additional isolation requirements were applied to either object.
The trigger efficiency for signal events passing the full selection is ${\sim}95\%$ and does not vary significantly with top squark mass or lifetime.

Events selected by the trigger are required to satisfy preselection criteria requiring a well-reconstructed electron and muon~\cite{Chatrchyan:2012xi, Chatrchyan:2013dga}.
Electron candidates are reconstructed by
combining information from deposits in the ECAL matched to tracks built by the
Gaussian sum filter algorithm~\cite{CMS-PAS-EGM-10-004,Chatrchyan:2013dga}.
Muon candidates are built by separately constructing a track from the muon path through the
tracking volume and one from its path through the muon systems, and then
performing a global fit between the two~\cite{Chatrchyan:2012xi}.
Jets are constructed using the implementation of the anti-\kt clustering algorithm available in
\textsc{FastJet}~\cite{FastJet} with a size parameter of 0.5~\cite{AntiKT} and are
required to have transverse momentum $\pt > 10\GeV$.
We require both leptons to have $\pt > 25\GeV$ and $\abs{\eta} < 2.5$,
where $\eta$ is the pseudorapidity,
and to pass the standard isolation requirements~\cite{Chatrchyan:2012xi, Chatrchyan:2013dga}.
As an additional isolation requirement,
there must not be any reconstructed jet within $\DR \equiv \sqrt{\smash[b]{(\Delta \eta)^2 + (\Delta \phi)^2}} < 0.5$ of either selected lepton,
where $\phi$ is the azimuthal angle of the momentum vector.
Electron candidates in the transition region between the barrel and the end cap detectors, corresponding to $1.44 < \abs{\eta} < 1.56$, are rejected.
Additionally, each lepton is required to have a transverse impact parameter ($d_0$) greater than $0.01\cm$,
where $d_0$ is defined as the distance of
closest approach in the transverse plane of the trajectory of the lepton track to the measured position of the center of the colliding proton beams.
Each event is required to have exactly one electron and one muon satisfying the criteria listed above.
The two leptons are required to be oppositely charged and separated by $\DR > 0.5$.

Events passing the preselection requirements are further categorized according to the lepton $d_0$.
When an unstable particle decays to charged particles,
the impact parameter values of the tracks associated with the decay products will be strongly correlated with the lifetime of the parent.
In order to be sensitive to a range of BSM particle lifetimes, we conduct this search in three nonoverlapping signal regions (SRn), defined in the 2D space of electron and muon $d_0$.
The SR3 region requires  $0.1\cm < d_0 < 2\cm$ for both the electron and the muon and is expected to have very low SM background contamination.
An event is classified in signal region SR2 if it fails the criteria of SR3 but both leptons have $d_0 > 0.05\cm$.
An event is classified in signal region SR1 if it fails the criteria of both SR3 and SR2 but both leptons have $d_0 > 0.02\cm$.
This last signal region is expected to contain more SM background, but has higher signal efficiency for the shortest BSM particle lifetimes to which this search is sensitive.
For simulated signal events, the selection efficiency is as high as $20\%$, for top squark lifetimes near $c\tau = 1\cm$.
This efficiency does not depend significantly on the top squark mass but is strongly dependent on top squark lifetime.

We use Monte Carlo simulation to estimate the signal acceptance as well as
contributions from several of the SM backgrounds.  Simulated samples of $\Pp\Pp
\to \sTop \sTop^*$ events were generated using \PYTHIA{}8~\cite{Pythia8}.
The samples simulating \cPZ+jets and single-top-quark production were generated using 
\POWHEG~\cite{Nason:2004rx,Frixione:2007vw,Alioli:2010xd,Alioli:2008gx,Alioli:2009je,Re:2010bp}, 
while those simulating \PW+jets and \ttbar were generated using \MADGRAPH
5~\cite{MADGRAPH}, with \PYTHIA{}6 used for hadronization for both generators~\cite{Sjostrand:2006za}.
Samples of diboson production were simulated with \PYTHIA{}6. For the
background samples generated with \POWHEG, the CT10 parton
distribution functions (PDF) are used~\cite{CT10}. All other samples use
CTEQ6L1~\cite{CTEQ6L1} PDFs.
All generated events have minimum bias interactions superimposed to simulate the effect of overlapping interactions within
the same event (pileup). They are then processed through a
detailed simulation of the CMS detector based on \GEANTfour~\cite{GEANT4}.
In the tables and figures that follow,
the ``Top quark'' sample is the combination of all \ttbar and single-top-quark
processes, the ``Other EW'' sample is the combination of \PW+jets, \zee, \zmumu, and
all diboson processes. The heavy-flavor QCD sample (``HF'') is obtained
directly from data, as described below.

We apply several correction factors to the Monte Carlo simulations to address known deficiencies in the modeling of data.
The simulation is adjusted to match the distribution of pileup observed in data.
Scale factors are derived to correct the performance of lepton identification and isolation algorithms.
The lepton corrections are estimated from data with the ``tag-and-probe" method~\cite{TagnProbe} using \zll events.
The simulated events are reweighted according to the data to simulation scale factors per lepton, as a function of $\pt$ and $\abs{\eta}$.
We use cosmic muon events in data and simulation to estimate the degradation of track reconstruction efficiency for large values of the impact parameter~\cite{DisplacedLeptons}.
The tracking efficiency is greater than 90\% in data and simulation and shows little variation for values of $d_0$ up to 2\cm.
We use these events to calculate a scale factor to correct the simulation to match the data in the region beyond $d_0 = 0.02\cm$.
This scale factor is measured to be $0.960 \pm 0.014$.
This factor is applied to leptons in simulated events that pass the selection criteria.
Trigger efficiencies are estimated using an independent data sample recorded with a combination of jet and \ETslash triggers.
This sample is enriched with \ttbar events by requiring the preselected events to
have an electron and a muon candidate, two jets with $\pt>30\GeV$ and $\abs{\eta} < 2.4$, and at least one b-tagged jet,
selected with the medium working point of the combined secondary vertex algorithm~\cite{CMS-PAS-BTV-13-001,Btagging}.
The ratio of efficiencies for data and simulated \ttbar events to pass the trigger requirements is used as a correction factor.
This scale factor is calculated to be $0.981 \pm 0.004$ and is applied to all the simulated data sets used in the search.

There are two SM sources of leptons with large displacements that constitute the dominant backgrounds to this search:
HF states, which can have lifetimes of $\tau \approx 0.05\cm/c$, and tau leptons ($\tau \approx 0.0087\cm/c$), which come mainly from \ztt decays.
The HF estimate is derived from the data-driven approach described below, while the \ztt estimate, as well as other subdominant background estimates, are taken from the simulation.

\begin{figure}[hbtp]
\centering
\includegraphics[width=0.44\textwidth]{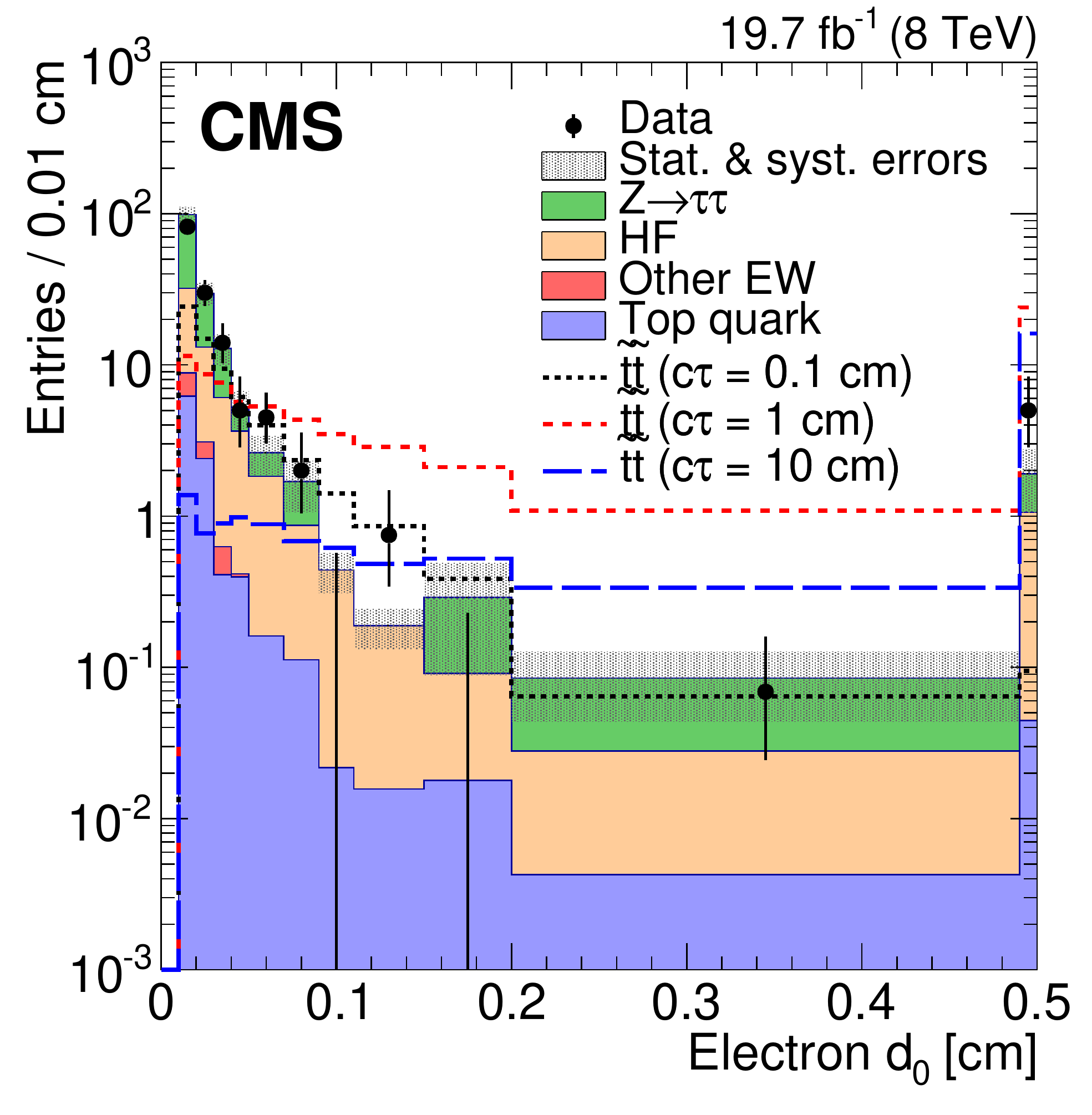}
\includegraphics[width=0.44\textwidth]{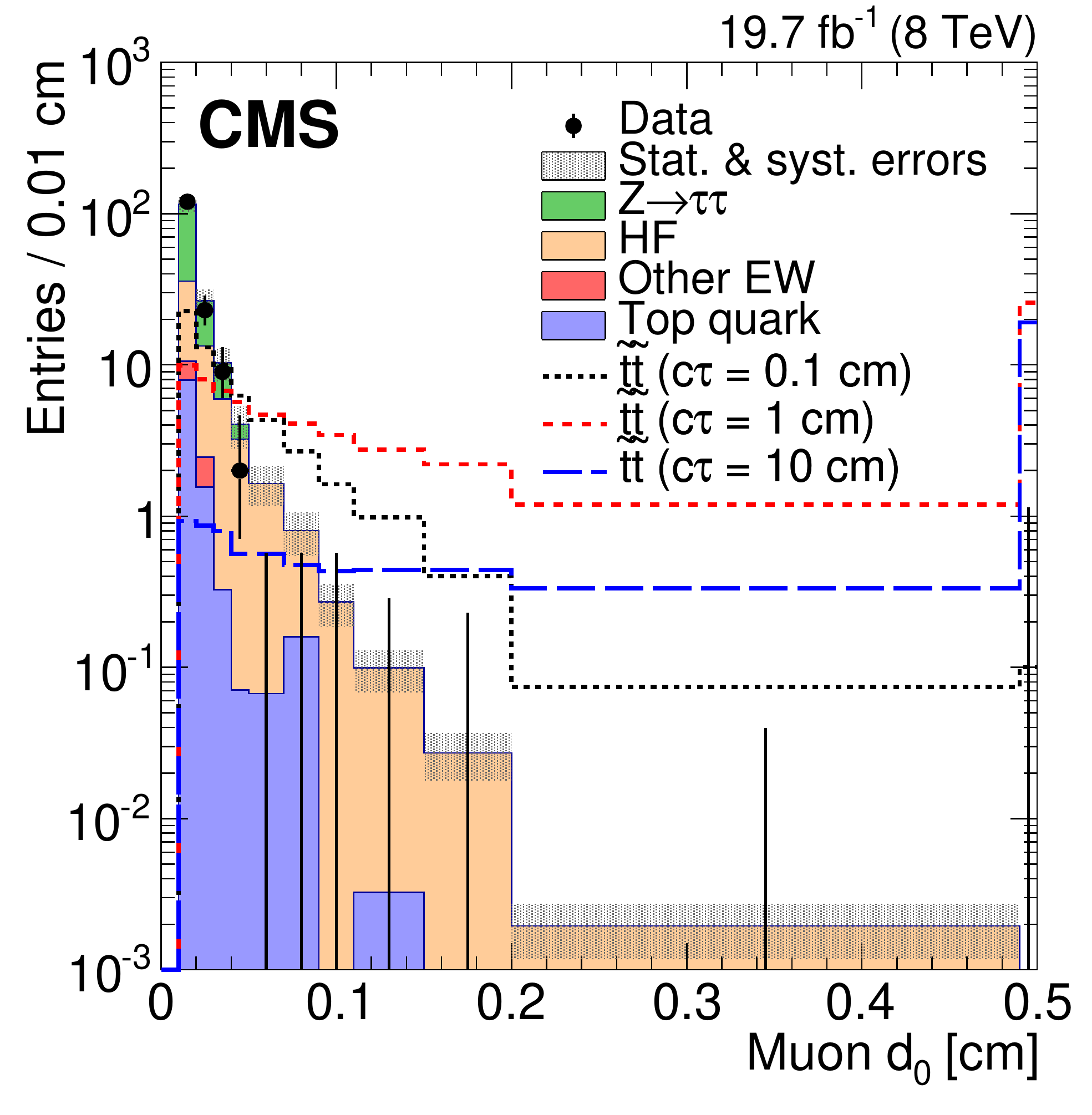}
\caption{Lepton transverse impact parameter distributions for data and expected
background processes after the preselection requirements have been applied, for
electrons (\cmsLeft) and muons (\cmsRight).  The signal expected from
500\GeV top squark pair production is overlaid for $c\tau = 0.1\cm$, 1\cm, and
10\cm. The event yields per bin have been rescaled to account for the varying
bin sizes.  The rightmost bin of each plot contains the overflow entries.}
\label{fig:displacedD0}
\end{figure}

The HF estimate is obtained using data collected on the same trigger described above.
A modified preselection is then applied, which excludes the requirements on lepton isolation and the sign of the lepton charge product.
We then use these two uncorrelated variables to define four non-overlapping regions.
We calculate the ratio of events in two of the four regions
and apply this ratio to the number of events in the third region to obtain an estimate of the number of HF events in the fourth region.
We define four regions corresponding to the following:
isolated leptons whose charges have the opposite sign (iso OS);
isolated leptons whose charges have the same sign (iso SS);
opposite sign leptons that are both anti-isolated (anti-iso OS);
and same sign leptons that are  both anti-isolated (anti-iso SS).
The iso OS region corresponds to the standard preselection criteria, while the other three regions are expected to contain primarily HF events.
The small contributions in these three regions from sources other than HF are subtracted using the simulation.
We then calculate the ratio of the number of HF events in the iso SS region
to the number of HF events in the anti-iso SS region.
We construct a HF background data set by taking events in the anti-iso OS region and normalizing them to the iso OS region using this ratio.
We use the impact parameter distributions from this data set, then apply the $d_0$ selections described above to obtain an estimate of the HF yield in the signal regions.
Figure~\ref{fig:displacedD0} shows the $d_0$ distributions for background and data after the preselection requirements.
The background prediction is $162 \pm 9\stat \pm 17\syst$ events and the observation is 154 events.

For each source of background that is estimated from simulation,
we calculate the efficiency for the selected electron and muon to pass the $d_0$ selections separately,
in order to overcome statistical limitations associated with sparsely populated regions of the simulated distributions.
We then estimate the number of events that satisfy both $d_0$ criteria by multiplying the two efficiencies
and applying this factor to the events that satisfy the preselection criteria.
In cases in which this method results in an estimated yield of zero events,
we use the yield calculated from the less exclusive neighboring search region,
which should always overestimate the background in that bin.
This technique produces virtually identical limits to simply using a null estimate in these cases.

There are several sources of systematic uncertainty in this search.
Cross section uncertainties for simulated data sets range from 4\% to 8\% for the SM background and 15\% to 28\% for the signal process.
Following the official PDF4LHC recommendation~\cite{Alekhin:2011sk}, we obtain PDF uncertainties using an envelope of several PDF sets, ranging from 1\% to 5\%.
The PDF and cross section uncertainties are propagated into the limits for all simulated data sets.
The luminosity estimate, based on the pixel cluster counting method~\cite{CMS-PAS-LUM-13-001}, has an uncertainty of 2.6\%.
For the displaced track reconstruction efficiency, the (4\%) correction is correlated for each lepton, resulting in a 8.0\% systematic uncertainty per event.
The uncertainty in the data-driven HF estimate is 30\% and is dominated by the limited size of the sample used.
Uncertainties in trigger efficiency, pileup correction, and lepton correction factors are calculated and incorporated but are small compared to the previously mentioned uncertainties.

\begin{table*}[tb]
\topcaption{Numbers of expected and observed events in the three search regions (see the text for the definitions of these regions).
Background and signal expectations are quoted as $N_{\text{exp}} \pm 1\sigma\stat \pm 1\sigma\syst$.
If the estimated background is zero in a particular search region, the estimate is instead taken from the preceding region.
Since this should always overestimate the background, we denote this by a preceding ``${<}$''.}
\renewcommand{\arraystretch}{1.2}
\resizebox{\textwidth}{!}{
\begin{scotch}{crrr}
Event source              & \multicolumn{1}{c}{SR1}     & \multicolumn{1}{c}{SR2}                   & \multicolumn{1}{c}{SR3} \\
\hline
Other\ EW                 & $0.65 \pm 0.13 \pm 0.09$    & $(0.89 \pm 0.53 \pm 0.12) \times 10^{-2}$ & ${<}(89 \pm 53 \pm 12) \times 10^{-4}$ \\
Top quark                 & $0.77 \pm 0.04 \pm 0.08$    & $(1.25 \pm 0.26 \pm 0.12) \times 10^{-2}$ & $(2.4 \pm 1.3 \pm 0.2) \times 10^{-4}$ \\
Z${\to}\tau\tau$  & $3.93 \pm 0.42 \pm 0.39$    & $(0.73 \pm 0.73 \pm 0.07) \times 10^{-2}$ & ${<}(73 \pm 73 \pm 7) \times 10^{-4}$ \\
HF                        & $12.7 \pm 0.2 \pm 3.8$      & $(98 \pm 6 \pm 30) \times 10^{-2}$        & $(340 \pm 110 \pm 100) \times 10^{-4}$ \\
\hline
Total expected background      & $18.0 \pm 0.5 \pm 3.8$      & $1.01 \pm 0.06 \pm 0.30$                  & $0.051 \pm 0.015 \pm 0.010$ \\
\hline
Observed                  & $19$                        & $0$                                       & $0$ \\
\hline
\hline
$\Pp\Pp {\to} \sTop \sTop^*$ ($M_{\sTop} = 500\GeV$) & & & \\
\hline
$c\tau = 0.1\cm$          & $30.1 \pm 0.7 \pm 5.3$      & $6.54 \pm 0.34 \pm 1.16$                  & $1.34 \pm 0.15 \pm 0.24$ \\
$c\tau = 1\cm$            & $35.3 \pm 0.8 \pm 6.2$      & $30.3 \pm 0.7 \pm 5.3$                    & $51.3 \pm 1.0 \pm 9.0$ \\
$c\tau = 10\cm$           & $4.73 \pm 0.30 \pm 0.83$    & $5.57 \pm 0.32 \pm 0.98$                  & $26.3 \pm 0.7 \pm 4.6$ \\
\end{scotch}
}
\label{tab:yields_Signal_Region}
\end{table*}

\begin{figure}[hbtp]
\centering
\includegraphics[width=0.48\textwidth]{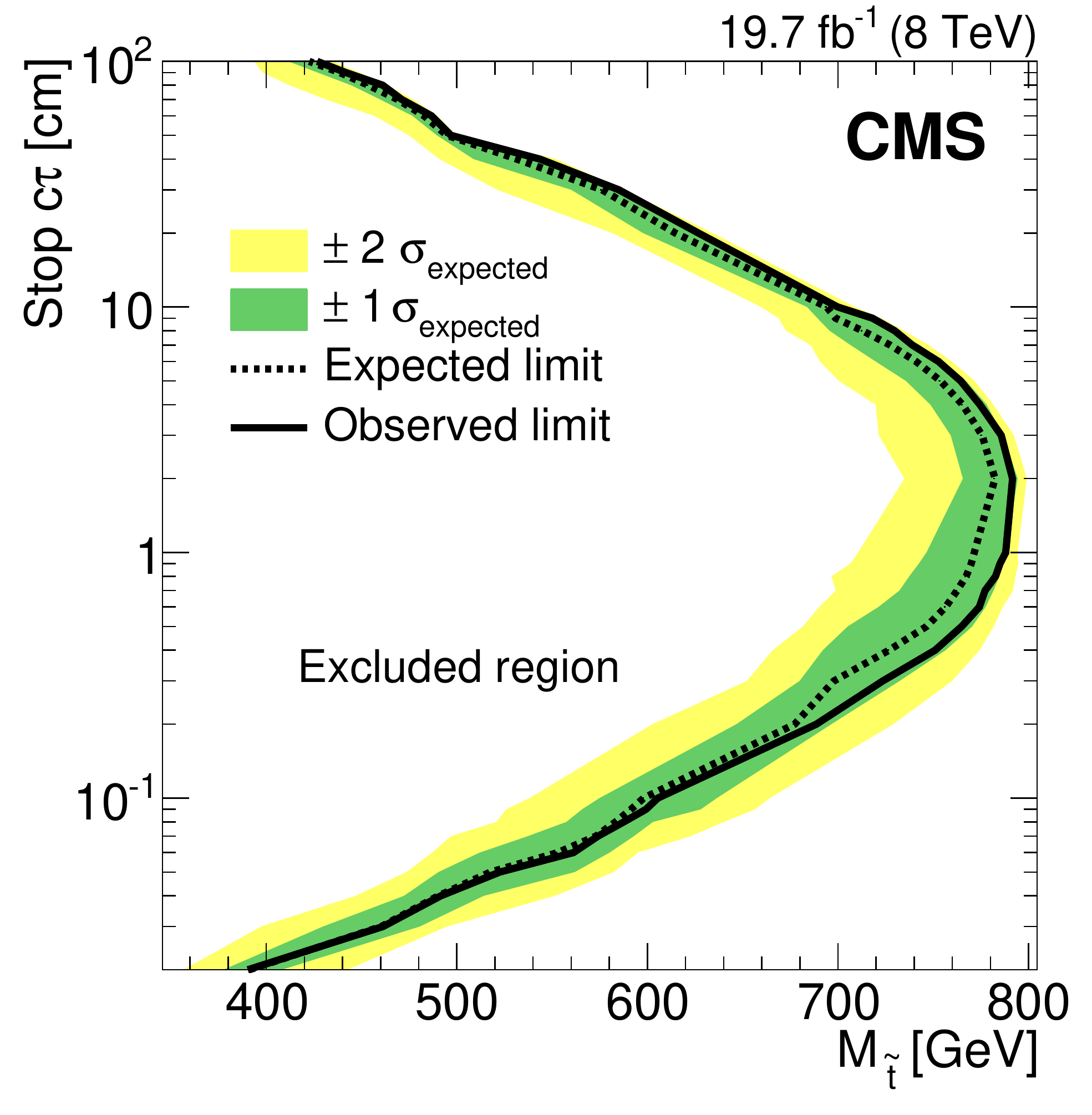}
\caption{Expected and observed 95\% \CL cross section exclusion contours for top squark pair production
in the plane of top squark lifetime ($c\tau$) and top squark mass.
These limits assume a branching fraction of 100\% through the RPV vertex
$\sTop \to \cPqb \Pl$, where the branching fraction to any lepton flavor is equal to 1/3.
As indicated in the plot, the region to the left of the contours is excluded by this search.}
\label{fig:ctauVsMass}
\end{figure}

Table~\ref{tab:yields_Signal_Region} shows the numbers of observed and expected background events in the three search regions.
We do not observe any significant excess over the background expectation.
We set 95\% confidence level (\CL) upper limits on the cross section for top squark pair production at 8\TeV.
We perform this as a simultaneous counting experiment in three bins of the three search regions.
We use a Bayesian calculation assuming a flat prior for the signal as a function of top squark mass.
Nuisance parameters arising from statistical uncertainties are modeled as gamma distributions,
while all others are modeled as log-normal distributions.
These cross section limits are translated into upper limits on the top squark mass,
where the cross section for each mass hypothesis is
calculated at next-to-leading-order and next-to-leading-logarithmic precision within a simplified model with decoupled
squarks and gluinos~\cite{SUSYCrossSections,SMS0,SMS1}.
The resulting expected and observed
limit contours are shown in Fig.~\ref{fig:ctauVsMass}. The region to the left of the contours is excluded.
For a lifetime of $c\tau = 2\cm$, we exclude top squark masses up to 790\GeV, to be compared with a value of 780\GeV expected in the absence of any signal.

In summary, a search has been performed for new physics with an electron and muon with opposite charges having a signature of large impact parameter values,
with no requirements made on jets or missing transverse energy.
No excess is observed above background for displacements up to 2\cm.
While this search is expected to have sensitivity to a wide range of theoretical models,
the results are interpreted in the context of a displaced supersymmetry model~\cite{DisplacedSUSY} with a pair-produced top squark having a lifetime between $c\tau = 0.02\cm$ and $c\tau = 100\cm$.
Limits are placed at 95\% C.L. on this model as a function of top squark mass and lifetime.
For a lifetime hypothesis of $c\tau = 2\cm$, top squark masses up to 790\GeV are excluded.
These are the most restrictive limits obtained to date on this model.

\begin{acknowledgments}
We congratulate our colleagues in the CERN accelerator departments for the excellent performance of the LHC and thank the technical and administrative staffs at CERN and at other CMS institutes for their contributions to the success of the CMS effort. In addition, we gratefully acknowledge the computing centers and personnel of the Worldwide LHC Computing Grid for delivering so effectively the computing infrastructure essential to our analyses. Finally, we acknowledge the enduring support for the construction and operation of the LHC and the CMS detector provided by the following funding agencies: BMWFW and FWF (Austria); FNRS and FWO (Belgium); CNPq, CAPES, FAPERJ, and FAPESP (Brazil); MES (Bulgaria); CERN; CAS, MoST, and NSFC (China); COLCIENCIAS (Colombia); MSES and CSF (Croatia); RPF (Cyprus); MoER, ERC IUT and ERDF (Estonia); Academy of Finland, MEC, and HIP (Finland); CEA and CNRS/IN2P3 (France); BMBF, DFG, and HGF (Germany); GSRT (Greece); OTKA and NIH (Hungary); DAE and DST (India); IPM (Iran); SFI (Ireland); INFN (Italy); NRF and WCU (Republic of Korea); LAS (Lithuania); MOE and UM (Malaysia); CINVESTAV, CONACYT, SEP, and UASLP-FAI (Mexico); MBIE (New Zealand); PAEC (Pakistan); MSHE and NSC (Poland); FCT (Portugal); JINR (Dubna); MON, RosAtom, RAS and RFBR (Russia); MESTD (Serbia); SEIDI and CPAN (Spain); Swiss Funding Agencies (Switzerland); MST (Taipei); ThEPCenter, IPST, STAR and NSTDA (Thailand); TUBITAK and TAEK (Turkey); NASU and SFFR (Ukraine); STFC (United Kingdom); DOE and NSF (U.S.).
\end{acknowledgments}

\bibliography{auto_generated}

\cleardoublepage \appendix\section{The CMS Collaboration \label{app:collab}}\begin{sloppypar}\hyphenpenalty=5000\widowpenalty=500\clubpenalty=5000\textbf{Yerevan Physics Institute,  Yerevan,  Armenia}\\*[0pt]
V.~Khachatryan, A.M.~Sirunyan, A.~Tumasyan
\vskip\cmsinstskip
\textbf{Institut f\"{u}r Hochenergiephysik der OeAW,  Wien,  Austria}\\*[0pt]
W.~Adam, T.~Bergauer, M.~Dragicevic, J.~Er\"{o}, M.~Friedl, R.~Fr\"{u}hwirth\cmsAuthorMark{1}, V.M.~Ghete, C.~Hartl, N.~H\"{o}rmann, J.~Hrubec, M.~Jeitler\cmsAuthorMark{1}, W.~Kiesenhofer, V.~Kn\"{u}nz, M.~Krammer\cmsAuthorMark{1}, I.~Kr\"{a}tschmer, D.~Liko, I.~Mikulec, D.~Rabady\cmsAuthorMark{2}, B.~Rahbaran, H.~Rohringer, R.~Sch\"{o}fbeck, J.~Strauss, A.~Taurok, W.~Treberer-Treberspurg, W.~Waltenberger, C.-E.~Wulz\cmsAuthorMark{1}
% ...
\vskip\cmsinstskip
% ...
1:~~Also at Vienna University of Technology, Vienna, Austria\\
2:~~Also at CERN, European Organization for Nuclear Research, Geneva, Switzerland\\
% ...

\end{sloppypar}
\end{document}